\documentclass{article}
\usepackage{graphicx}
\usepackage{amsmath}

\begin{document}

\noindent \textbf{Jackson's paradox and its resolution by the
four-dimensional }

\noindent \textbf{geometric quantities }\medskip \bigskip

\qquad Tomislav Ivezi\'{c}

\qquad\textit{Ru%
\mbox
{\it{d}\hspace{-.15em}\rule[1.25ex]{.2em}{.04ex}\hspace{-.05em}}er Bo\v
{s}kovi\'{c} Institute, P.O.B. 180, 10002 Zagreb, Croatia}

\textit{\qquad ivezic@irb.hr\bigskip \bigskip }

\noindent In this paper it is shown that the real cause of Jackson's paradox
is the use of three-dimensional (3D) quantities, e.g., $\mathbf{E}$, $%
\mathbf{B}$, $\mathbf{F}$, $\mathbf{L}$, $\mathbf{T}$, their transformations
and equations with them. The principle of relativity is naturally satisfied
and there is no paradox when the physical reality is attributed to the 4D
geometric quantities, e.g., to the 4D torque $N$ (bivector) or,
equivalently, to the 4D torques $N_{s}$ and $N_{t}$ (1-vectors), which
together contain the same physical information as the bivector $N$. \bigskip
\medskip

\noindent \textbf{I. INTRODUCTION\bigskip }

In a recent paper$^{1}$ in this Journal Jackson discovered an apparent
paradox; there is a three-dimensional (3D) torque and so a time rate of
change of 3D angular momentum in one inertial frame, but no 3D angular
momentum and no 3D torque in another. Two inertial frames $S$ (the
laboratory frame) and $S^{\prime }$ (the moving frame) are considered (they
are $K$ and $K^{\prime }$ respectively in Jackson's notation). In $S^{\prime
}$ a particle of charge $q$ and mass $m$ experiences only the radially
directed electric force caused by a point charge $Q$ fixed permanently at
the origin. Consequently both $\mathbf{L}^{\prime }$ and the torque $\mathbf{%
T}^{\prime }$ (Jackson's $\mathbf{N}^{\prime }$ is denoted as $\mathbf{T}%
^{\prime }$) are zero in $S^{\prime }$, see Fig. 1(a) in Ref. 1. (The
vectors in the 3D space will be designated in bold-face.) In $S$ the charge $%
Q$ is in uniform motion and it produces \emph{both} an electric field $%
\mathbf{E}$ and a \emph{magnetic field} $\mathbf{B}$. The existence of $%
\mathbf{B}$ in $S$ is responsible for the existence of the 3D magnetic force
$\mathbf{F}=q\mathbf{u}\times \mathbf{B}$ and this force provides a 3D
torque $\mathbf{T}$ ($\mathbf{T}=\mathbf{x}\times \mathbf{F}$) on the
charged particle, see Fig. 1(b) in Ref. 1. Consequently a nonvanishing
angular momentum of the charged particle changes in time in $S$, $\mathbf{T}%
=d\mathbf{L}/dt$. Jackson$^{1}$ considers that there is no paradox and that
such result is relativistically correct result, i.e., that the principle of
relativity is not violated.

It is recently revealed$^{2}$ that the real cause of the paradox is - the
use of 3D quantities, e.g., $\mathbf{E}$, $\mathbf{B}$, $\mathbf{F}$, $%
\mathbf{L}$, $\mathbf{T}$, their transformations and equations with them. In
Ref. 2, instead of using 3D quantities, it is dealt from the outset with 4D
geometric quantities and equations with them. In such treatment the paradox
does not appear and the principle of relativity is naturally satisfied. In
this paper we shall briefly repeat the consideration from Ref. 2. It is
worth noting that exactly the same paradox appears in the Trouton-Noble
experiment, see, e.g., Ref. 3 and references therein.

The calculations in Ref. 2 (also in Ref. 3) and in this paper are performed
in the geometric algebra formalism, which is recently nicely presented in
this Journal by Hestenes.$^{4}$ Physical quantities will be represented by
geometric 4D quantities, multivectors, that are defined without reference
frames, i.e., as absolute quantities (AQs) or, when some basis has been
introduced, they are represented as 4D coordinate-based geometric quantities
(CBGQs) comprising both components and a basis. For simplicity and for
easier understanding only the standard basis \{$\gamma _{\mu };0,1,2,3$\} of
orthonormal 1-vectors, with timelike vector $\gamma _{0}$ in the forward
light cone, will be used. For all mathematical details regarding the
spacetime algebra reader can consult Hestenes' paper.$^{4}$\textit{\bigskip
\bigskip }

\noindent \textbf{II. DISCUSSION\ OF\ JACKSON'S DERIVATION\ AND\ RESULTS\
\bigskip }

First a remark about the figures in Ref. 1. Both figures would need to
contain the time axes as well. Fig. 1(a) (Fig. 1(b), and Fig. 2) is the
projection onto the hypersurface $t^{\prime }=const.$ ($t=const.$); the
distances are simultaneously determined in the $S^{\prime }$ ($S$) frame.
The Lorentz transformations (LT) cannot transform the hypersurface $%
t^{\prime }=const.$ into the hypersurface $t=const.$ and the distances that
are simultaneously determined in the $S^{\prime }$ frame cannot be
transformed by the LT into the distances simultaneously determined in the $S$
frame. Such figures could be possible for the Galilean transformations, but
for the LT they are meaningless.

In Sec. III Jackson$^{1}$ discusses ``Lorentz transformations of the angular
momentum between frames.'' He starts with the usual covariant definition of
the angular momentum tensor $M^{\mu \nu }=x^{\mu }p^{\nu }-x^{\nu }p^{\mu }$%
, Eq. (8) in Ref. 1. Notice that the standard basis $\left\{ \gamma _{\mu
}\right\} $, i.e., Einstein's system of coordinates, is implicit in that
definition. In Einstein's system of coordinates the standard, i.e.,
Einstein's synchronization$^{5}$ of distant clocks and Cartesian space
coordinates $x^{i}$ are used in the chosen inertial frame. In Ref. 1 the
vector\emph{\ }$\mathbf{L}$ is called the angular momentum and its
components $L_{i}$ are identified with the space-space components of $M^{\mu
\nu }$. However, a physical interpretation is not given for another vector $%
\mathbf{L}_{t}$. Its components $L_{t,i}$ are identified with the three
time-space components of $M^{\mu \nu }$ (we denote Jackson's $K_{i}$ with $%
L_{t,i}$, $\mathbf{K}$ with $\mathbf{L}_{t}$). The same identification is
supposed to hold both in $S^{\prime }$, the rest frame of the charges $q$
and $Q$, $L_{i}^{\prime }=(1/2)\varepsilon _{ikl}M^{\prime kl}$, $%
L_{t,i}^{\prime }=M^{\prime 0i}$, and in $S$, the laboratory frame; the same
relations but with unprimed quantities. This then leads to the usual
transformations of the components of $\mathbf{L}$ that are given by Eq. (11)
in Ref. 1. In contrast to Ref. 1 we write both transformations, for $L_{i}$
and for $L_{t,i}$; $L_{1}=L_{1}^{\prime }$, $L_{2}=\gamma (L_{2}^{\prime
}-\beta L_{t,3}^{\prime })$, $L_{3}=\gamma (L_{3}^{\prime }+\beta
L_{t,2}^{\prime })$, and $L_{t,1}=L_{t,1}^{\prime }$, $L_{t,2}=\gamma
(L_{t,2}^{\prime }+\beta L_{3}^{\prime })$, $L_{t,3}=\gamma (L_{t,3}^{\prime
}-\beta L_{2}^{\prime })$. The characteristic feature of these
transformations is that the components $L_{i}$ in $S$ are expressed by the
mixture of components $L_{i}^{\prime }$ and $L_{t,i}^{\prime }$ from $%
S^{\prime }$. This causes that the components of the 3D angular momentum do
not vanish in the laboratory frame $S$, even if they do in $S^{\prime }$. In
the case considered in Ref. 1 $L_{3}$ is different from zero due to
contribution from $L_{t,2}^{\prime }$.

This is in a complete analogy with the usual identification of the
components of $\mathbf{B}$ and $\mathbf{E}$ with the space-space and the
time-space components respectively of the electromagnetic field strength
tensor $F^{\mu \nu }$. In $S^{\prime }$ it is $B_{i}^{\prime
}=(1/2c)\varepsilon _{ikl}F^{\prime lk}$, $E_{i}^{\prime }=F^{\prime i0}$,
and the same relations but with unprimed quantities hold in $S$, see Ref. 6
Sec. 11.9. Such procedure yields the usual transformations for the
components of $\mathbf{B}$ and $\mathbf{E}$, see, e.g., Ref. 6 Eq. (11.148).
The comparison of the identifications shows that the components $L_{i}$
correspond to $-B_{i}$ and $L_{t,i}$ to $-E_{i}$. In all these relations the
components of the 3D vectors $\mathbf{L}$, $\mathbf{L}_{t}$ and $\mathbf{B}$%
, $\mathbf{E}$ are written with lowered (generic) subscripts, since they are
not the spatial components of the 4D quantities. This refers to the
third-rank antisymmetric $\varepsilon $ tensor too. The super- and
subscripts are used only on the components of the 4D quantities.

In the same way we can determine the components $T_{i}$ of the torque $%
\mathbf{T}$ ($\mathbf{T}=d\mathbf{L}/dt$) and the components $T_{t,i}$ of
another torque $\mathbf{T}_{t}\ $($\mathbf{T}_{t}=d\mathbf{L}_{t}/dt$) by
the identification in both frames with the space-space and the time-space
components respectively of the torque four-tensor $N^{\mu \nu }$, $%
T_{i}=(1/2)\varepsilon _{ikl}N^{kl}$, $T_{t,i}=T^{0i}$, which gives the
transformations
\begin{eqnarray}
T_{1} &=&T_{1}^{\prime },\ T_{2}=\gamma (T_{2}^{\prime }-\beta
T_{t,3}^{\prime }),\ T_{3}=\gamma (T_{3}^{\prime }+\beta T_{t,2}^{\prime }),
\notag \\
T_{t,1} &=&T_{t,1}^{\prime },\ T_{t,2}=\gamma (T_{t,2}^{\prime }+\beta
T_{3}^{\prime }),\ T_{t,3}=\gamma (T_{t,3}^{\prime }-\beta T_{2}^{\prime }).
\label{tc}
\end{eqnarray}
Again the transformed components $T_{i}$ are expressed by the mixture of
components $T_{k}^{\prime }$ and $T_{t,k}^{\prime }$. Hence the components
of ``physical'' torque $\mathbf{T}$ do not vanish in $S$, even if they do in
$S^{\prime }$. In Ref. 1 all $T_{k}^{\prime }$ are zero but the component $%
T_{3}$ in $S$ is $\neq 0$ due to contribution from $T_{t,2}^{\prime }$.
Jackson,$^{1,6}$ as all others, considers that only the space-space parts, $%
L_{i}$ and $T_{i}$, i.e. the 3D vectors $\mathbf{L}$ and $\mathbf{T}$, are
physical quantities. Actually, it is almost generally accepted that the
covariant quantities, e.g., $M^{\mu \nu }$, $N^{\mu \nu }$, $F^{\mu \nu }$,
etc. are only auxiliary mathematical quantities from which ``physical'' 3D
quantities, $\mathbf{L}$, $\mathbf{T}$, $\mathbf{E}$ and $\mathbf{B}$, etc.,
are deduced.

Several objections are raised in Refs. 2, 7-9 to such derivation of the
usual transformations for the components of $\mathbf{L}$ and $\mathbf{L}_{t}$%
, $\mathbf{B}$ and $\mathbf{E}$ and $\mathbf{T}$ and $\mathbf{T}_{t}$. Some
of them are quoted here.

(i) $M^{\mu \nu }$, $F^{\mu \nu }$ and $N^{\mu \nu }$ are only components
(numbers) that are (implicitly) determined in the standard basis $\left\{
\gamma _{\mu }\right\} $, i.e., in Einstein's system of coordinates. In
another system of coordinates that is different than the Einstein system of
coordinates, e.g., differing in the chosen synchronization, all above
identifications are impossible and meaningless.

(ii) The above identifications and transformations refer only to components.
The 3D vectors, e.g., $\mathbf{T}^{\prime }$ and $\mathbf{T}_{t}^{\prime }$
in $S^{\prime }$, and $\mathbf{T}$ and $\mathbf{T}_{t}$ in $S$, which are
geometric quantities in the 3D space, are constructed multiplying six
independent components of $N^{\prime \mu \nu }$ and $N^{\mu \nu }$ by the
unit 3D vectors $\mathbf{i}^{\prime }$, $\mathbf{j}^{\prime }$, $\mathbf{k}%
^{\prime }$ in $S^{\prime }$ and $\mathbf{i}$, $\mathbf{j}$, $\mathbf{k}$ in
$S$. The components, e.g., $T_{i}$ in $S$ are determined by the usual
transformations (\ref{tc}) from the components $T_{i}^{\prime }$ and $%
T_{t,i}^{\prime }$ in $S^{\prime }$, but there is no transformation which
transforms $\mathbf{i}^{\prime }$, $\mathbf{j}^{\prime }$, $\mathbf{k}%
^{\prime }$ from $S^{\prime }$ into $\mathbf{i}$, $\mathbf{j}$, $\mathbf{k}$
in $S$. Hence it is not true that $\mathbf{T}=T_{1}\mathbf{i}+T_{2}\mathbf{j}%
+T_{3}\mathbf{k}$ is obtained by the LT from $\mathbf{T}^{\prime }\mathbf{=}%
T_{1}^{\prime }\mathbf{i}^{\prime }+T_{2}^{\prime }\mathbf{j}^{\prime
}+T_{3}^{\prime }\mathbf{k}^{\prime }$. Consequently $\mathbf{T}$ and $%
\mathbf{T}^{\prime }$ are not the same quantity for relatively moving
inertial observers in $S$ and $S^{\prime }$, $\mathbf{T\neq T}^{\prime }$.
The transformations that do not refer to the same 4D quantity are not the
LT, but we call them the ``apparent'' transformations (AT). According to
that the transformations (\ref{tc}) and those for $\mathbf{L}$, Eq. (11) in
Ref. 1, are not the LT, but they are the AT of the 3D $\mathbf{T}$ and $%
\mathbf{L}$. In Ref. 7-9 it is explained in detail that the usual
transformations of $\mathbf{B}$ and $\mathbf{E}$, Ref. 6 Eqs. (11.148) and
(11.149), are also the AT and not the LT (a fundamental achievement). In
fact, as shown in Ref. 2, all transformations of the 3D vectors $\mathbf{p}$%
, $\mathbf{F}$,\textbf{\ }$\mathbf{E}$, $\mathbf{B}$, $\mathbf{L}$, $\mathbf{%
T}$, etc. are the AT and not the LT. This also means that equations with
them, like Eq. (4) ($d\mathbf{p}/dt=\mathbf{F}$ $=q\mathbf{E}+q\mathbf{u}%
\times \mathbf{B}$) and Eq. (5) ($d\mathbf{L}/dt=\mathbf{T}$) in Ref. 1, are
not relativistically correct equations. Thus, the fact that in Ref. 1 the
same result, $\mathbf{T}^{\prime }=\mathbf{0}$ but $\mathbf{T}\neq \mathbf{0}
$, is obtained by two different methods of calculation does not mean that
the calculations are relativistically correct and that the principle of
relativity is satisfied. \textit{\bigskip \bigskip }

\noindent \textbf{III. DEFINITIONS\ OF\ 4D\ QUANTITIES \bigskip }

The relativistically correct treatment of the problem considered in Ref. 1
and the resolution of Jakson's paradox are given in Ref. 2, Secs. 4-4.3,
using 4D torques $N$, $N_{s}$ and $N_{t}$. Here we shall only briefly expose
the consideration from Secs. 2, 4, 4.1 and 4.3, Ref. 2.

As shown in Ref. 10 the electromagnetic field $F$ (bivector) is the primary
quantity for the whole electromagnetism. $F(x)$ for a charge $Q$ moving with
constant velocity $u_{Q}$ (1-vector) is
\begin{equation}
F(x)=kQ(x\wedge (u_{Q}/c))/\left| x\wedge (u_{Q}/c)\right| ^{3},  \label{cvf}
\end{equation}
where $k=1/4\pi \varepsilon _{0}$. For the charge $Q$ at rest, $%
u_{Q}/c=\gamma _{0}$. Instead of the usual identification of $E_{i}$ and $%
B_{i}$ with components of $F^{\mu \nu }$ we construct, in a mathematically
correct way, the 4D geometric quantities that represent the electric and
magnetic fields by a decomposition of $F$. It can be decomposed into
1-vectors of the electric field $E$ and the magnetic field $B$ and a unit
time-like 1-vector $v/c$ as
\begin{align}
F& =(1/c)E\wedge v+(IB)\cdot v,  \notag \\
E& =(1/c)F\cdot v,\quad B=-(1/c^{2})I(F\wedge v),  \label{itf}
\end{align}
see Ref. 8. In (\ref{itf}) $I$ is the unit pseudoscalar. ($I$ is defined
algebraically without introducing any reference frame, as in Ref. 11, Sec.
1.2.) $v$ is the velocity (1-vector) of a family of observers who measures $%
E $ and $B$ fields. Since $F$ is antisymmetric it holds that $E\cdot
v=B\cdot v=0$, which yields that only three components of $E$ and three
components of $B$ are independent quantities. Observe that $E$ and $B$
depend not only on $F$ but on $v$ as well.

From (\ref{itf}) and the known $F$, Eq. (\ref{cvf}), we find the 1-vectors $%
E $ and $B$ for a charge $Q$ moving with constant velocity $u_{Q}$
\begin{eqnarray}
E &=&(D/c^{2})[(x\wedge u_{Q})\cdot v]  \notag \\
B &=&(-D/c^{3})I(x\wedge u_{Q}\wedge v),  \label{ec}
\end{eqnarray}
where $D=kQ/\left| x\wedge (u_{Q}/c)\right| ^{3}$. Note that $B$ in (\ref{ec}%
) can be expressed in terms of $E$ as $B=(1/c^{3})I(u_{Q}\wedge E\wedge v).$
When the world lines of the observer and the charge $Q$ coincide, $u_{Q}=v$,
then (\ref{ec}) yields that $B=0$ and only an electric field (Coulomb field)
remains. $E$ and $B$ from (\ref{ec}) are AQs; they are independent of the
chosen reference frame and the chosen system of coordinates in it. They
naturally generalize the expressions for the 3D $\mathbf{E}$ and $\mathbf{B}$%
, Eqs. (3a) and (3b) in Ref. 1, to the 4D spacetime.

Furthermore, instead of the covariant form of the Lorentz force, Eq. (A2) in
Ref. 1 (only components in the implicit $\{\gamma _{\mu }\}$ basis), we deal
with the Lorentz force as an AQ, $K_{L}=(q/c)F\cdot u$. Using the
decomposition of $F$ into $E$ and $B$, Eq. (\ref{itf}), this Lorentz force
can be written as $K_{L}=(q/c)\left[ (1/c)E\wedge v+(IB)\cdot v\right] \cdot
u$, where $u$ is the velocity (1-vector) of a charge $q$. Particularly, from
the definition of the Lorentz force $K_{L}=(q/c)F\cdot u$ and the relation $%
E=(1/c)F\cdot v$ it follows that the Lorentz force ascribed by an observer
comoving with a charge, $u=v$, is purely electric $K_{L}=qE$. Hence our
equation of motion of a charge $q$ is $mdu/d\tau =K_{L}=(q/c)\left[
(1/c)E\wedge v+(IB)\cdot v\right] \cdot u$, which replaces Eq. (4) from Ref.
1. In that equation $\tau $ is the proper time, $u$ is the velocity 1-vector
of a particle that is defined to be the tangent to its world line.

The 4D AQs, the angular momentum $M$ and the torque $N$ (bivectors) for the
Lorentz force $K_{L}$ and manifestly Lorentz invariant equation connecting $%
M $ and $N$ are defined as
\begin{equation}
M=x\wedge p,\quad N=x\wedge K_{L};\quad N=dM/d\tau ,  \label{MKN}
\end{equation}
where $x$ is the position 1-vector and $p$ is the proper momentum (1-vector)
$p=mu$.

When $M$ and $N$ are written as CBGQs in the $\{\gamma _{\mu }\}$ basis they
become $M=(1/2)M^{\mu \nu }\gamma _{\mu }\wedge \gamma _{\nu }$, with $%
M^{\mu \nu }=x^{\mu }p^{\nu }-x^{\nu }p^{\mu }$, and $N=(1/2)N^{\mu \nu
}\gamma _{\mu }\wedge \gamma _{\nu }$, with $N^{\mu \nu }=x^{\mu }K_{L}^{\nu
}-x^{\nu }K_{L}^{\mu }$. The components, e.g., $N^{\mu \nu }$, are
determined as $N^{\mu \nu }=\gamma ^{\nu }\cdot (\gamma ^{\mu }\cdot
N)=(\gamma ^{\nu }\wedge \gamma ^{\mu })\cdot N$. We see that the components
$M^{\mu \nu }$ are identical to the covariant angular momentum four-tensor
given by Eq. (A3) in Ref. 1. However $M$ and $N$ from (\ref{MKN}) are 4D
geometric quantities, the 4D AQs, whereas the components $M^{\mu \nu }$ and $%
N^{\mu \nu }$ that are used in the usual covariant approach, e.g., Eq. (A3)
in Ref. 1, are coordinate quantities, the numbers obtained in the specific
system of coordinates, i.e., in the $\{\gamma _{\mu }\}$ basis. In contrast
to the usual covariant approach, $M$ and $N$ as 4D CBGQs are also 4D
geometric quantities, which contain both components and a \emph{basis}, here
bivector basis $\gamma _{\mu }\wedge \gamma _{\nu }$. The essential
difference between our geometric approach and the usual covariant picture is
the presence of the basis. \emph{The existence of a basis causes that every
4D CBGQ is invariant under the passive LT}; the components transform by the
LT and the basis by the inverse LT leaving the whole 4D CBGQ unchanged. This
means that a CBGQ represents \emph{the same physical quantity }for
relatively moving 4D observers. Hence it holds that, e.g.,
\begin{equation}
N=(1/2)N^{^{\prime }\mu \nu }\gamma _{\mu }^{\prime }\wedge \gamma _{\nu
}^{\prime }=(1/2)N^{\mu \nu }\gamma _{\mu }\wedge \gamma _{\nu },  \label{nj}
\end{equation}
where all primed quantities are the Lorentz transforms of the unprimed ones.
When physical laws are written with such Lorentz invariant quantities, 4D
AQs or 4D CBGQs, then they automatically satisfy the principle of
relativity. In the standard approach to special relativity$^{5}$ the
principle of relativity is postulated outside the framework of a
mathematical formulation of the theory. There,$^{5}$ it is also considered
that the principle of relativity holds for the equations written with the 3D
quantities.

Here, as in Ref. 2, we introduce new 4D torques as AQs, 1-vectors $N_{s}$
and $N_{t}$. The same decomposition can be made for $N$ as for $F$, (\ref
{itf}). It is decomposed into two 1-vectors, the ``space-space'' torque $%
N_{s}$ and the ``time-space'' torque $N_{t}$, and the unit time-like
1-vector $v/c$ as
\begin{eqnarray}
N &=&(v/c)\cdot (IN_{s})+(v/c)\wedge N_{t}  \notag \\
N_{s} &=&I(N\wedge v/c),\ N_{t}=(v/c)\cdot N;\ N_{s}\cdot v=N_{t}\cdot v=0.
\label{nls}
\end{eqnarray}
Only three components of $N_{s}$ and three components of $N_{t}$ are
independent since $N$ is antisymmetric. Here again $v$ is the velocity
(1-vector) of a family of observers who measures $N_{s}$ and $N_{t}$. Again,
as for $E$ and $B$, the torques $N_{s}$ and $N_{t}$ depend not only on the
bivector $N$ but on $v$ as well. The relations (\ref{nls}) show that $N_{s}$
and $N_{t}$ taken\emph{\ together} contain the same physical information as
the bivector $N$.

When $N_{s}$ and $N_{t}$ are written as CBGQs in the $\{\gamma _{\mu }\}$
basis they are $N_{s}=N_{s}^{\mu }\gamma _{\mu }=(1/2c)\varepsilon ^{\alpha
\beta \mu \nu }N_{\alpha \beta }v_{\mu }\gamma _{\nu }$, $N_{t}=(1/c)N^{\mu
\nu }v_{\mu }\gamma _{\nu }$. In the frame of ``fiducial'' observers (it
will be called the $\gamma _{0}$-frame), in which the observers who measure
1-vectors $E$, $B$, $K_{L}$, $N_{s}$ and $N_{t}$ are at rest, the velocity $%
v $ is $v=c\gamma _{0}$. In that frame and in the $\{\gamma _{\mu }\}$ basis
$v^{\mu }=(c,0,0,0)$. Let us now take that the $S$ frame is the $\gamma _{0}$%
-frame. Then in $S$, $N_{s}^{0}=N_{t}^{0}=0$, and only the spatial
components remain. $N_{s}^{i}$ components are $%
N_{s}^{1}=N^{23}=x^{2}K_{L}^{3}-x^{3}K_{L}^{2}$, $N_{s}^{2}=N^{31}$ and $%
N_{s}^{3}=N^{12}$. Comparison with the identification $T_{i}=(1/2)%
\varepsilon _{ikl}N^{kl}$ shows that in the $\gamma _{0}$-frame the
components of the 1-vector $N_{s}$ correspond to components of ``physical''
3D torque $\mathbf{T}$, $T_{i}=N_{s}^{i}$, and similarly $T_{t,i}=N_{t}^{i}$%
. Hence only for ``fiducial'' observers one can deal with \emph{components}
of \emph{two} 3D torques $\mathbf{T}$ and $\mathbf{T}_{t}$; all \emph{six}
components are equally well physical. However, even in that frame the
relativistically correct \emph{geometric} quantities are not 3D vectors $%
\mathbf{T}$ and $\mathbf{T}_{t}$, but 1-vectors $N_{s}$ and $N_{t}$. The
whole discussion with the torque can be completely repeated for the angular
momentum replacing $N$, $N_{s}$ and $N_{t}$ by $M$, $M_{s}$ and $M_{t}$. In
the $\gamma _{0}$-frame components of 1-vectors $M_{s}$ and $M_{t}$
correspond to components of $\mathbf{L}$ and $\mathbf{L}_{t}$ respectively
in the usual 3D picture.

Of course, it holds that, e.g., $N_{s}$ as a CBGQ\ is a Lorentz invariant
quantity, $N_{s}=N_{s}^{\mu }\gamma _{\mu }=N_{s}^{\prime \mu }\gamma _{\mu
}^{\prime }$, where again all primed quantities are the Lorentz transforms
of the unprimed ones. The components of $N_{s}$ ($N_{t}$, $M_{s}$, $M_{t}$, $%
E$, $B$, ..) transform under the LT as the components of any 1-vector
transform
\begin{equation}
N_{s}^{\prime 0}=\gamma (N_{s}^{0}-\beta N_{s}^{1}),\ N_{s}^{\prime
1}=\gamma (N_{s}^{1}-\beta N_{s}^{0}),\ N_{s}^{\prime 2,3}=N_{s}^{2,3}.
\label{ans}
\end{equation}
The LT of $N_{t}^{\mu }$, $M_{s}^{\mu }$, $M_{t}^{\mu }$ and of $E^{\mu }$, $%
B^{\mu }$ are of the same form. In contrast to all mentioned AT for the
components of the 3D vectors, e.g., Eq. (\ref{tc}) or Eq. (11) in Ref. 1 for
$L_{i}$, the components $N_{s}^{\mu }$ transform again to $N_{s}^{\prime \mu
}$; \emph{there is no mixing of components}.

Furthermore we write $N$ from (\ref{MKN}) using the expression for $K_{L}$
and Eq. (\ref{cvf}) for $F$,
\begin{equation}
N=(Dq/c^{2})(u\cdot x)(u_{Q}\wedge x).  \label{AN}
\end{equation}
$N_{s}$ and $N_{t}$ are then determined from (\ref{nls}) and (\ref{AN})
\begin{eqnarray}
N_{s} &=&(Dq/c^{3})(u\cdot x)I(x\wedge v\wedge u_{Q}),  \notag \\
N_{t} &=&(Dq/c^{3})(u\cdot x)[(x\wedge u_{Q})\cdot v].  \label{n1}
\end{eqnarray}
Comparison with (\ref{ec}) shows that $N_{s}$ and $N_{t}$ can be expressed
in terms of $B$ and $E$ as
\begin{eqnarray}
N_{s} &=&q(u\cdot x)B,  \notag \\
N_{t} &=&(q/c)(u\cdot x)E.  \label{n2}
\end{eqnarray}
As already said, in connection with (\ref{ec}), when $u_{Q}=v$ then $B=0$
and $N_{s}=0$ as well.\medskip \bigskip

\noindent \textbf{IV. RESOLUTION\ OF\ JACKSON'S\ PARADOX USING}

\textbf{4D TORQUES \bigskip }

The knowledge of $N$ as an AQ, Eq. (\ref{AN}), enables us to find the
expressions for $N$ as CBGQs in $S^{\prime }$ and $S$. First let us write
all AQs from (\ref{AN}) as CBGQs in $S^{\prime }$, the rest frame of the
charge $Q$, in which $u_{Q}=c\gamma _{0}^{\prime }$. Then $N=(Dq/c)(u\cdot
x)(\gamma _{0}^{\prime }\wedge x)$, and in the $\{\gamma _{\mu }^{\prime }\}$
basis it is explicitly given as
\begin{eqnarray}
N &=&(1/2)N^{\prime \mu \nu }\gamma _{\mu }^{\prime }\wedge \gamma _{\nu
}^{\prime }=N^{\prime 01}(\gamma _{0}^{\prime }\wedge \gamma _{1}^{\prime
})+N^{\prime 02}(\gamma _{0}^{\prime }\wedge \gamma _{2}^{\prime }),  \notag
\\
N^{\prime 01} &=&(Dq/c)(u^{\prime \mu }x_{\mu }^{\prime })x^{\prime 1},\
N^{\prime 02}=(Dq/c)(u^{\prime \mu }x_{\mu }^{\prime })x^{\prime 2}.
\label{nc}
\end{eqnarray}
The components $x^{\prime \mu }$ are $x^{\prime \mu }=(x^{\prime
0}=ct^{\prime },x^{\prime 1},x^{\prime 2},0)$ where $x^{\prime 1}=r^{\prime
}\cos \theta ^{\prime }$, $x^{\prime 2}=r^{\prime }\sin \theta ^{\prime }$.
In $S^{\prime }$, $u=u^{\prime \mu }\gamma _{\mu }^{\prime }$, where $%
u^{\prime \mu }=dx^{\prime \mu }/d\tau =(u^{\prime 0},u^{\prime 1},u^{\prime
2},0)$. The components $N^{^{\prime }\mu \nu }$ that are different from zero
are only $N^{\prime 01}$ and $N^{\prime 02}$.

In order to find the torque $N$ in $S$ we now write all AQs from (\ref{AN})
as CBGQs in $S$ and in the $\{\gamma _{\mu }\}$ basis. In $S$ the charge $Q$
is moving with velocity $u_{Q}=\gamma _{Q}c\gamma _{0}+\gamma _{Q}\beta
_{Q}c\gamma _{1}$, where $\beta _{Q}=\left| \mathbf{u}_{Q}\right| /c$ and $%
\gamma _{Q}=(1-\beta _{Q}^{2})^{-1/2}$. Then $N$ is
\begin{equation}
N=(1/2)N^{\mu \nu }\gamma _{\mu }\wedge \gamma _{\nu }=N^{01}\gamma
_{0}\wedge \gamma _{1}+N^{02}\gamma _{0}\wedge \gamma _{2}+N^{12}\gamma
_{1}\wedge \gamma _{2},  \label{g}
\end{equation}
or explicitly it becomes
\begin{eqnarray}
N &=&(Dq/c)(u^{\mu }x_{\mu })[\gamma _{Q}(x^{1}-\beta _{Q}x^{0})(\gamma
_{0}\wedge \gamma _{1})  \notag \\
&&+\gamma _{Q}x^{2}(\gamma _{0}\wedge \gamma _{2})+\beta _{Q}\gamma
_{Q}x^{2}(\gamma _{1}\wedge \gamma _{2})].  \label{n}
\end{eqnarray}
Now the components $N^{\mu \nu }$ that are different from zero are not only
the time-space components $N^{01}$ and $N^{02}$ but also the space-space
component $N^{12}=\beta _{Q}N^{02}$.

We note that another way to find $N$ as CBGQ in $S$, (\ref{n}), is to make
the LT of $N$ as CBGQ in $S^{\prime }$, (\ref{nc}). Of course, according to (%
\ref{nj}), both CBGQs are equal, $N$ ((\ref{nc})) $=$ $N$ ((\ref{n})); they
represent the same 4D quantity $N$ from (\ref{AN}) in $S^{\prime }$ and $S$
frames. The principle of relativity is naturally satisfied and there is not
any paradox.

Instead of the torque $N$ we can use $N_{s}$ and $N_{t}$. However, as
already said, $N_{s}$ and $N_{t}$ are not uniquely determined by $N$, but
their explicit values depend also on $v$. This means that it is important to
know which frame is chosen to be the $\gamma _{0}$-frame. Observe that the
same conclusions also refer to the determination of $E$, $B$ and $M_{s}$, $%
M_{t}$ from $F$ and $M$ respectively. In this paper we shall only consider
the case when $S$ is the $\gamma _{0}$-frame. For another case readers can
consult Sec. 4.2 in Ref. 2. $N_{s}$ and $N_{t}$ will be determined directly
from (\ref{n1}) taking into account that $v=c\gamma _{0}$ and $u_{Q}=\gamma
_{Q}c\gamma _{0}+\gamma _{Q}\beta _{Q}c\gamma _{1}$. This yields that
\begin{equation}
N_{s}=N_{s}^{\mu }\gamma _{\mu }=N^{12}\gamma _{3},\ N_{t}=N_{t}^{1}\gamma
_{1}+N_{t}^{2}\gamma _{2}=N^{01}\gamma _{1}+N^{02}\gamma _{2},  \label{2n}
\end{equation}
where $N^{12}$, $N^{01}$ and $N^{02}$ are from (\ref{n}). Thus when $S$ is
the $\gamma _{0}$-frame \emph{the ``space-space'' torque }$N_{s}$ \emph{is
different from zero. }

The same $N_{s}$ and $N_{t}$ can be determined using (\ref{n2}) and (\ref{ec}%
). The charge\textbf{\ }$Q$ moves in $S$, which yields that both $E$ and
\emph{the magnetic field} $B$\emph{\ are different from zero.} Then $%
E=E^{\mu }\gamma _{\mu }$, $E^{0}=E^{3}=0$, $E^{1}=D\gamma _{Q}(x^{1}-\beta
_{Q}x^{0})$, $E^{2}=D\gamma _{Q}x^{2}$, and the magnetic field is $B=B^{\mu
}\gamma _{\mu }$, $B^{0}=B^{1}=B^{2}=0$, $B^{3}=(D/c)\gamma _{Q}\beta
_{Q}x^{2}=\beta _{Q}E^{2}/c$. The spatial components $E^{i}$ and $B^{i}$ are
the same as the usual expressions for the components of $\mathbf{E}$ and $%
\mathbf{B}$ for an uniformly moving charge. Inserting these equations into (%
\ref{n2}) we again find $N_{s}$ and $N_{t}$ as in (\ref{2n}). $N_{s}$ is $%
\neq 0$ since $B$ is $\neq 0$.

$N_{s}$ from (\ref{2n}) can be written in the form similar to Eq. (7) from
Ref. 1, when the above explicit form of $E$ and $B$ are used. Thus
\begin{equation}
N_{s}=N^{3}\gamma _{3}=N^{12}\gamma _{3}=(\beta _{Q}ctK_{L}^{2}+(q/c)\beta
_{Q}y(E^{\mu }u_{\mu })\gamma _{3}.  \label{nw}
\end{equation}
In the usual approach, e.g., Ref. 1, it is considered that in the $S$ frame
the whole physical torque is the 3D $\mathbf{T}$, i.e., $T_{z}$, given by
Eq. (7) in Ref. 1. We see that in the 4D spacetime the physical torque,
theoretically and \emph{experimentally, }is either the bivector $N$, (\ref{n}%
) or (\ref{nc}), or \emph{two} 1-vectors $N_{s}$ and $N_{t}$ given by (\ref
{2n}) and (\ref{nw}). Only when the laboratory frame $S$ is the $\gamma _{0}$%
-frame the spatial components of $N_{s}$ can be put into the correspondence
with the components of the 3D $\mathbf{T}$. However note that in (\ref{nw})
all components are the components of the 4D quantities, 1-vectors $x,$ $u,$ $%
K_{L},$ $E$ and $B$, while in Eq. (7) in Ref. 1 only the corresponding 3D
vectors are involved.

Let us now determine $N_{s}$ and $N_{t}$ as CBGQs in $S^{\prime }$. Relative
to the $S^{\prime }$ frame the charge $Q$ is at rest $u_{Q}=c\gamma
_{0}^{\prime }$, but the ``fiducial'' observers are moving with velocity $%
v=\gamma _{Q}c\gamma _{0}^{\prime }-\gamma _{Q}\beta _{Q}c\gamma
_{1}^{\prime }$. Then $N_{s}$ and $N_{t}$ in $S^{\prime }$ can be obtained
either directly from (\ref{n1}) or by means of the LT (\ref{ans}) of $N_{s}$
and $N_{t}$ from (\ref{2n}). We find $N_{s}$ and $N_{t}$ as
\begin{eqnarray}
N_{s} &=&N_{s}^{\prime \mu }\gamma _{\mu }^{\prime }=N_{s}^{\prime 3}\gamma
_{3}^{\prime },\ N_{s}^{\prime 3}=\gamma _{Q}\beta _{Q}N^{\prime 02},  \notag
\\
N_{t} &=&N_{t}^{\prime \mu }\gamma _{\mu }^{\prime },\ N_{t}^{\prime
0}=-\beta _{Q}\gamma _{Q}N^{\prime 01},\ N_{t}^{\prime 1,2}=\gamma
_{Q}N^{\prime 01,2},\ N_{t}^{\prime 3}=0,  \label{s}
\end{eqnarray}
where $N^{\prime 01}$, $N^{\prime 02}$ are given in (\ref{nc}). Now $N_{s}$
is different from zero not only in $S$ but in the $S^{\prime }$ frame as
well. The same results for $N_{s}$ and $N_{t}$ in $S^{\prime }$ can be
obtained using (\ref{n2}) and (\ref{ec}) and writing all AQs as CBGQs in the
$S^{\prime }$ frame.

It can be easily seen that $N_{s}$ ($N_{t}$) from (\ref{2n}) is equal to $%
N_{s}$ ($N_{t}$) from (\ref{s}); it is the same 4D CBGQ for observers in $S$
and $S^{\prime }$; the principle of relativity is naturally satisfied and
there is no paradox.

Inserting $N_{s}$ and $N_{t}$ from (\ref{2n}) and (\ref{s}) into Eq. (\ref
{nls}), which connects $N$ with $N_{s}$ and $N_{t}$, we find that the
expressions for $N$ in $S$ and $S^{\prime }$ are the same, as it must be.

If we would take that $S^{\prime }$ is the $\gamma _{0}$-frame, as in Sec.
4.2 in Ref. 2, then the explicit expressions for $N_{s}$ and $N_{t}$ as
CBGQs would be different than those given in (\ref{2n}) and (\ref{s}). For
example, in that case $N_{s}$ is zero but $N_{t}\neq 0$ both in $S^{\prime }$
and $S$. However when these new expressions for $N_{s}$ and $N_{t}$ as CBGQs
are inserted into (\ref{nls}) they will give the same $N$ as when $S$ is the
$\gamma _{0}$-frame. \bigskip \medskip

\noindent \textbf{V. CONCLUSIONS\bigskip }

It is proved in this paper, as in Ref. 2, that the physical torques are the
4D geometric quantities, the bivector $N$ defined in (\ref{MKN}) and (\ref
{AN}), or the 1-vectors $N_{s}$ and $N_{t}$ that are derived from $N$
according to (\ref{nls}). They together contain the same physical
information as the bivector $N$. In the considered case $N_{s}$ and $N_{t}$
are defined in (\ref{n1}). Only in the $\gamma _{0}$-frame one can deal with
components $T_{i}$ and $T_{t,i}$ of two 3D torques $\mathbf{T}$ and $\mathbf{%
T}_{t}$ respectively. In that frame temporal components of $N_{s}$ and $%
N_{t} $ as CBGQs are zero, $N_{s}^{0}=N_{t}^{0}=0$, and the spatial
components are $T_{i}=N_{s}^{i}$ and $T_{t,i}=N_{t}^{i}$. All components $%
T_{i}$ and $T_{t,i}$ are equally well physical for ``fiducial'' observers.
Hence, it is not true, as generally accepted, that only the 3D torque $%
\mathbf{T}$ is a well-defined physical quantity. However, it is shown here,
and in Ref. 2, that even in the frame of ``fiducial'' observers the
relativistically correct geometric quantities are not 3D vectors $\mathbf{T}$
and $\mathbf{T}_{t}$, but the bivector $N$ or 1-vectors $N_{s}$ and $N_{t}$.
These 4D geometric quantities correctly transform under the LT, e.g., the
whole torque $N_{s}$ remains unchanged under the passive LT, $%
N_{s}=N_{s}^{\mu }\gamma _{\mu }=N_{s}^{\prime \mu }\gamma _{\mu }^{\prime }$%
, where the components transform by means of the LT (\ref{ans}) and the
basis 1-vectors $\gamma _{\mu }$ by the inverse LT. This means that the
principle of relativity is satisfied and the paradox with the torque does
not appear. In contrast to it the components $T_{i}$ and $T_{t,i}$ transform
according to the AT (\ref{tc}), which differ from the LT for components of
1-vectors, e.g., (\ref{ans}). Furthermore, the objections (i) and (ii) from
Sec. II show that the transformations of $\mathbf{T}$ and $\mathbf{T}_{t}$,
as geometric quantities in the 3D space, are not the LT but the
relativistically incorrect AT.

The validity of the above relations with 4D geometric quantities can be
experimentally checked measuring all six independent components of $N$, or $%
N_{s}$ and $N_{t}$ taken together, in both relatively moving frames. Only
such complete data are physically relevant in the 4D spacetime. Remember
that the usual 3D torque $\mathbf{T}$ is connected only with three spatial
components of $N_{s}$ in the frame of ``fiducial'' observers. These three
components are not enough for the determination of the relativistically
correct 4D torques $N$, or $N_{s}$ and $N_{t}$. This is the real cause of
Jackson's paradox. \bigskip \medskip

\noindent \textbf{REFERENCES\bigskip }

\noindent $^{1}$J.D. Jackson, ``Torque or no torque? Simple charged particle
motion observed

in different inertial frames,'' Am. J. Phys. \textbf{72,} 1484-1487 (2004).

\noindent $^{2}$T. Ivezi\'{c}, ``Four-dimensional geometric quantities
versus the usual three-dimensional

quantities: The resolution of Jackson's paradox,'' Found. Phys. \textbf{36},
(10)

(2006), published, Issue: Online First; physics/0602105.

\noindent $^{3}$T. Ivezi\'{c}, ``Trouton-Noble paradox revisited,''
physics/0606176

\noindent $^{4}$D. Hestenes, ``Spacetime physics with geometric algebra,''
Am. J Phys. \textbf{71},

691-714 (2003).

\noindent $^{5}$A. Einstein, ``Zur Elektrodynamik bewegter K\"{o}rper,''
Ann. Physik. (Leipzig),

\textbf{17}, 891-921 (1905), tr. by W. Perrett and G.B. Jeffery, in \textit{%
The Principle of}

\textit{Relativity,} (Dover, New York, 1952).

\noindent $^{6}$J.D. Jackson, \textit{Classical Electrodynamics} (Wiley, New
York, 1977) 2nd ed.

\noindent $^{7}$T. Ivezi\'{c}, ``The proof that the standard transformations
of E and B are not

the Lorentz transformations'' Found. Phys. \textbf{33}, 1339-1347 (2003)%
\textbf{.}

\noindent $^{8}$T. Ivezi\'{c}, ``The difference between the standard and the
Lorentz

transformations of the electric and magnetic fields. Application to motional

EMF,'' Found. Phys. Lett. \textbf{18,} 301-324 (2005).

\noindent $^{9}$T. Ivezi\'{c}, ``The proof that Maxwell equations with the
3D E and B are not

covariant upon the Lorentz transformations but upon the standard

transformations: The new Lorentz invariant field equations ,''

Found. Phys. \textbf{35,} 1585-1615 \textbf{(}2005\textbf{)}.

\noindent $^{10}$T. Ivezi\'{c}, ``Axiomatic geometric formulation of
electromagnetism with only

one axiom: The field equation for the bivector field F with an explanation

of the Trouton-Noble experiment,'' Found. Phys. Lett. \textbf{18,} 401-429
(2005)\textbf{.}

\noindent $^{11}$D. Hestenes and G. Sobczyk, \textit{Clifford Algebra to
Geometric Calculus}

(Reidel, Dordrecht, 1984).

\end{document}